\documentclass[11pt,preprint]{aastex}
\usepackage{lscape}
\usepackage{epsfig}
\usepackage{ulem}
\usepackage{soul}

\begin{document}

\title{The KELT-South Telescope\footnote{The KELT-South Telescope is funded and operated by Vanderbilt University and Fisk University in cooperation with the South African Astronomical Observatory.}}
\author
{Joshua Pepper\altaffilmark{1}, Rudolf B. Kuhn\altaffilmark{2,3}, Robert Siverd\altaffilmark{1}, David James\altaffilmark{4}, Keivan Stassun\altaffilmark{1,5}
}
\altaffiltext{1}{Department of Physics \& Astronomy, Vanderbilt University, 6301 Stevenson Center, Nashville, TN 37235}
\altaffiltext{2}{South African Astronomical Observatory, P.O. Box 9, Observatory 7935, Cape Town, South Africa}
\altaffiltext{3}{Astrophysics, Cosmology and Gravity Centre, Department of Astronomy, University of Cape Town, South Africa}
\altaffiltext{4}{Cerro Tololo InterAmerican Observatory, Colina El Pino, S/N, La Serena, Chile}
\altaffiltext{5}{Department of Physics, Fisk University, 1000 17th Ave. N., Nashville, TN 37208}

\bibliographystyle{apj}

\begin{abstract}
The Kilodegree Extremely Little Telescope (KELT) project is a survey for new transiting planets around bright stars.  KELT-South is a small-aperture, wide-field automated telescope located at Sutherland, South Africa. The telescope surveys a set of $26^{\circ} \times 26^{\circ}$ fields around the southern sky, and targets stars in the range of $8 < V < 10$ mag, searching for transits by Hot Jupiters. This paper describes the KELT-South system hardware and software and discusses the quality of the observations. We show that KELT-South is able to achieve the necessary photometric precision to detect transits of Hot Jupiters around solar-type main-sequence stars.
\end{abstract}

\maketitle


\section{Introduction}

Ground-based surveys for transiting exoplanets have become increasingly successful, identifying nearly 130 transiting planets\footnote{http://exoplanet.eu/}.  These surveys, notably HATNet \citep{Bakos:2004}, SuperWASP \citep{Pollacco:2006}, and XO \citep{McCullough:2005}, utilize small-aperture, wide-field robotic telescopes to obtain high-precision photometric lightcurves of relatively bright stars.

The Kilodegree Extremely Little Telescope (KELT) project follows a similar methodology, originally based on the model derived in \citet{pgd}, with the KELT-North telescope \citep{Pepper:2007} operating for several years at Winer Observatory\footnote{http://www.winer.org/} in Arizona.  While KELT-North has been gathering data, we have constructed a duplicate telescope for Southern hemisphere observations: KELT-South.  The KELT-South telescope was deployed to the Sutherland observing station of the South African Astronomical Observatory (SAAO) in 2009, and is now operating regularly, accumulating data for transit detection and other science.  In this paper, we describe our hardware and facilities (\S \ref{sec:HF}) and observing operations and data handling procedures (\S \ref{sec:obs_data}).  We then characterize the telescope performance (\S \ref{sec:char}) and image quality (\S \ref{sec:image}), and conclude with sample lightcurves of interesting variable objects (\S \ref{sec:eg}).

\section{Hardware and Facilities}\label{sec:HF}

\begin{figure}
  \epsscale{0.75}
  \begin{center}
    \plotone{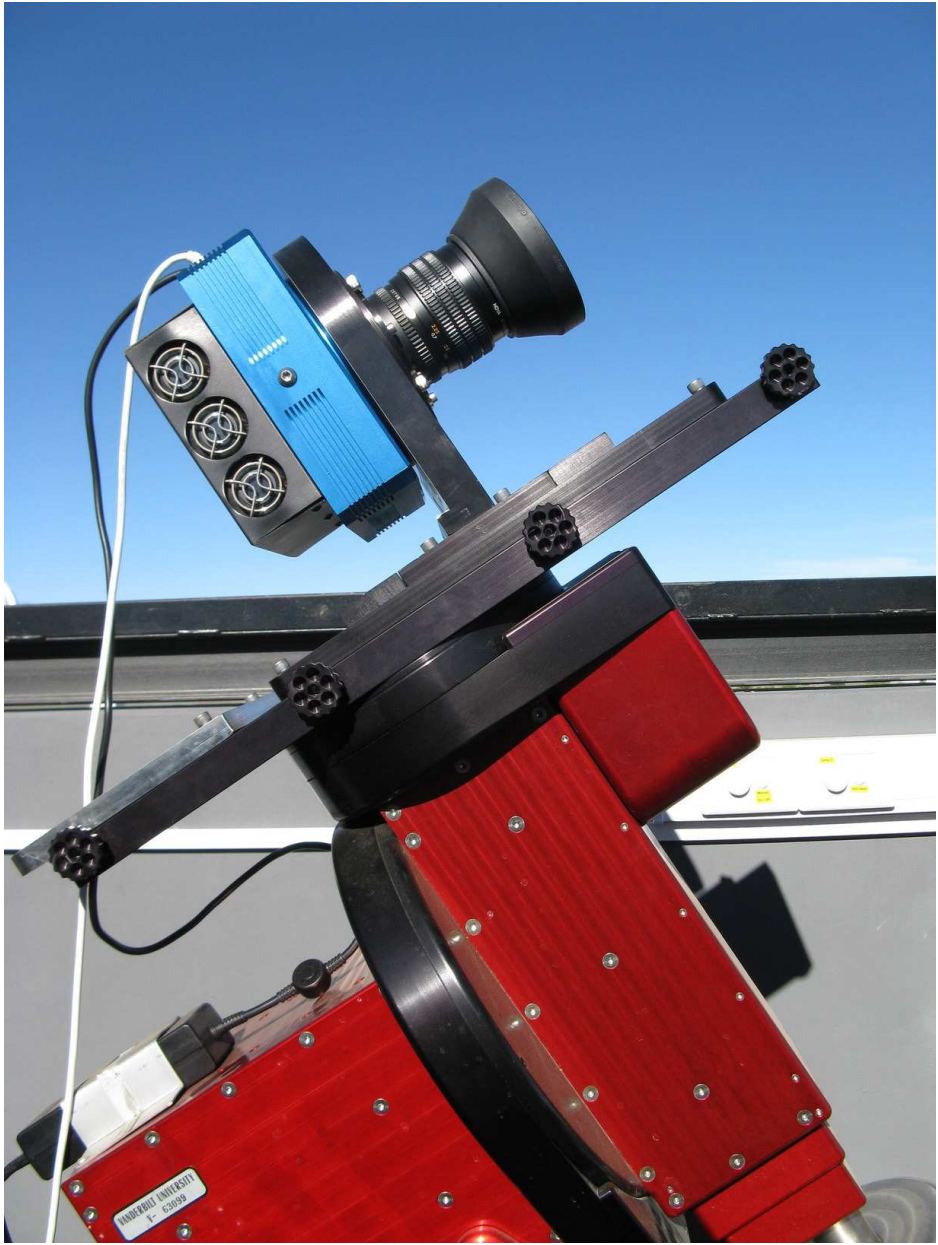}
    \caption{The KELT-South telescope installed at Sutherland.}
    \label{fig:pic}
  \end{center}
\end{figure}

KELT-South consists of an optical assembly (CCD Detector, camera lens, and filter) mounted on a robotic telescope mount (see Figure \ref{fig:pic}).  The mount is housed inside an enclosure with a roll-off roof that was specifically built for the telescope. Inside the enclosure there is a climate-controlled cabinet that houses the control computer and hard drives.  The control computer is responsible for controlling the robotic mount, CCD camera, observing operations, preliminary image processing and local data archiving. KELT-South was assembled using as many off-the-shelf components as possible to speed up the development process, as well as to ensure that replacing any part of the telescope would be relatively easy and quick.

\epsscale{1.0}

\subsection{Enclosure and PLC} \label{sec:Enc_PLC}
The KELT-South telescope is located in a custom-built building to protect it from the elements as well as hold the various system components. The building is a brick construction with a steel roll-off roof, equipped with power and network connections.  Cowlings on either side of the building allow air to flow through the building and keep components inside near the ambient temperature.  A flat field screen is bolted to the underside of the roll-off roof and interior lights can be turned on remotely to check on the condition of the telescope via an IP camera attached to the inside of the building.

Control of the building mechanisms is mediated by a custom-built programmable logic controller (PLC) located in a box on the interior wall of the building, remotely accessible through the control computer.  The PLC is connected to an uninterruptible power supply (UPS) that is responsible for closing the roof in the case of a power failure (see Section \ref{sec:UPS}).  The PLC has a hard-coded trigger that will close the roof and switch off any lights when no commands have been received from the control computer for a set amount of time.  This is known as the ``watchdog'' and is currently set to 900 seconds.  This ensures that the enclosure is closed and set to an ``off'' state in the event that the control computer fails or crashes during an observing run when the roof is open.

The control computer monitors the weather via three local weather station feeds provided by the SALT (Southern African Large Telescope)\footnote{http://www.salt.ac.za/} weather station\footnote{http://www.salt.ac.za/html/weather.php}, the SuperWASP\footnote{http://www.superwasp.org/waspsouth.htm} weather station \footnote{http://wasp.astro.keele.ac.uk/live/}, and the GFZ\footnote{http://www.gfz-potsdam.de/} weather station.  An external rain sensor is attached to the KELT-South building and serves as an extra safety feature.  This rain sensor is connected directly to the PLC and will close the roof in the case of rain detection while the roof is open.  Should the weather information collected by the control computer be wrong or the control computer not respond fast enough to incoming precipitation, the rain sensor will trigger and close the roof to prevent damage to the telescope and other systems.  The rain sensor is located on the side of the building that faces west, the direction from which storms typically approach the Sutherland observing site.

\subsection{Control Computer and Cabinet}
\label{sec:CtrlComp}
The computer that controls all aspects of the telescope operation is a Dell Optiplex 755 small-form-factor PC running the Windows XP operating system.  Software packages (TheSky6 Professional\footnote{http://www.bisque.com/sc/pages/thesky6family.aspx} [Version 6.0.0.60] and CCDSoft\footnote{http://www.bisque.com/sc/shops/store/CCDSoftWin2.aspx} [Version 5.00.188]) provided by Software Bisque\footnote{http://www.bisque.com/sc/}, enable the control computer to operate the CCD and mount via a script-accessible interface.  The computer clock is synchronized using the free software package Dimension 4\footnote{http://www.thinkman.com/dimension4/} to keep the clock time accurate. Two external hard drives are connected to the control computer via USB cables and are used to store the data. 

The control computer is housed in a Rittal computer cabinet (type 8845.500) with a Rittal Top Therm Plus Cooling Unit.  The cabinet also contains a transformer (\S \ref{sec:UPS})  to convert the local South African power (220V, 50Hz) to American standards (120V, 60Hz), required for the the American-manufactured components (i.e. camera, mount, and computer).  The cooling unit is set up to keep the interior temperature below 20$^\circ$C to protect the electronic components.

\subsection{Temperature Probes}
\label{sec:TempProbes}
Two temperature probes record the temperatures inside and outside the computer cabinet. These probes are connected to the control computer through a USB interface, and software on the computer logs the temperatures at one minute intervals. The probes were manufactured by Qti (Quality Thermistor, inc)\footnote{http://www.thermistor.com/productsDirecTemp.html} and provide absolute accuracy of up to $\pm$0.1$^\circ$C.

At present the temperatures are simply recorded and no actions are taken if the temperatures are outside the control limits. In the future, the control computer will alert the technical staff in Sutherland if this situation should happen.

\subsection{UPS and Transformer}
\label{sec:UPS}
The telescope operates with two UPS units and a transformer.  The first UPS is an Eaton Powerware 9125 UPS. Attached to this UPS is an Eaton PowerPass Distribution Module transformer.  Together the two systems are responsible for providing surge protection to the telescope as well as stepping down the power from 220V (South African standard) to 120V (American standard).  The 120V output power is required to operate the mount, camera and control computer (all these components were bought in the USA).  The second UPS is an Eaton Powerware 9120 UPS that provides backup power to the enclosure only. The roof motor, interior lights (including flat field lights) and internal IP camera are connected to this UPS.  The PLC is plugged into this UPS as well as a standard power outlet. If the normal power should fail, the PLC will sense the failure and will close the roof of the enclosure within 120 seconds if normal power is not restored.  This action cannot be stopped and is hardwired into the circuitry of the PLC.  This prevents the roof from staying open during a power failure and limits the possibility of damage to the telescope.  If this action should occur, the PLC trips an electronic flag that needs to be reset remotely by a user to open the roof again.

\subsection{Camera, Lens, and Filter}
\label{sec:Camera}
\begin{figure}[ht]
	\begin{center}
		\plotone{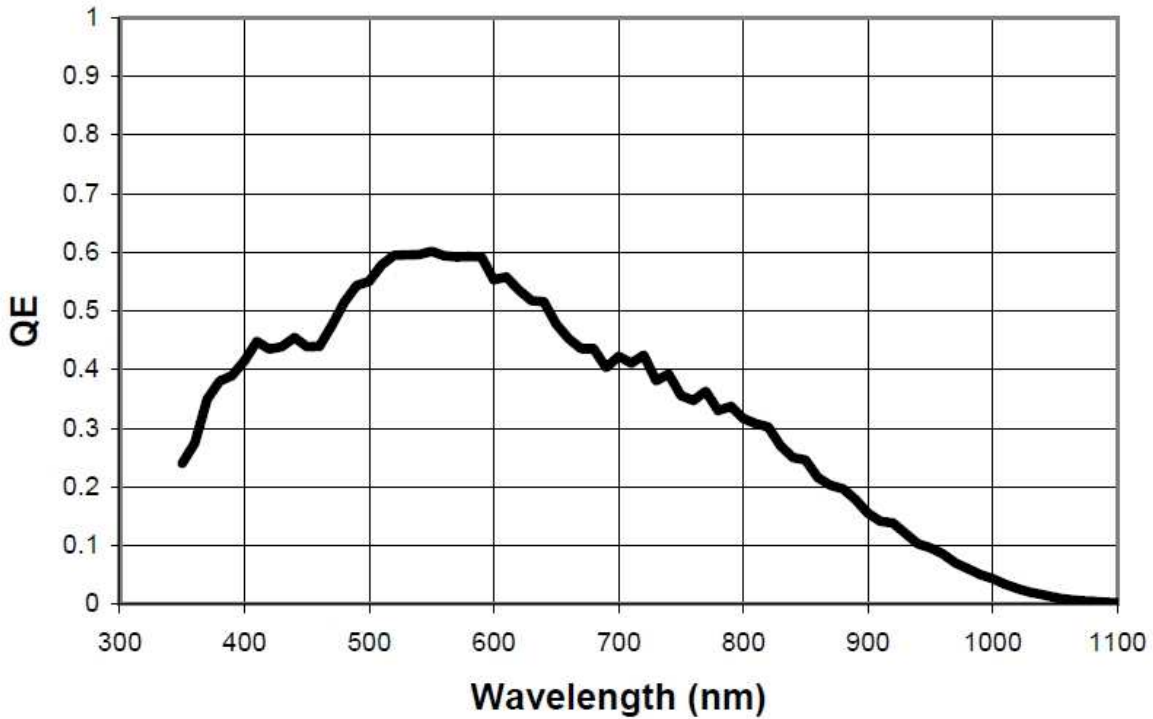}
		\caption{The theoretical response function of the KELT-South CCD as provided by Kodak.}
		\label{fig:QEimage}
	\end{center}
\end{figure}

The KELT-South detector is an Apogee Instruments Alta U16M thermoelectrically cooled CCD camera. The camera uses the Kodak KAF-16803 front illuminated CCD with 4096 $\times$ 4096 9$\mu$m pixels (36.88 $\times$ 36.88 $mm^2$ detector area) and has a peak quantum efficiency of $\sim$60\% at 550nm\footnote{http://www.kodak.com/global/en/business/ISS/Products/Fullframe/index.jhtml?pq-path=11937/11938/14425} (See Figure \ref{fig:QEimage}). The interface between the camera and the computer is a 5m long USB 2.0 standard cable. The camera has a three-stage thermoelectric Peltier cooler (with the D09F Apogee camera housing) that is capable of maintaining a temperature of 65 -- 70$^\circ$C below ambient temperature.  For the purpose of KELT-South the CCD is maintained at -20$^\circ$C year-round to reduce thermal noise.  The device is read out with 16-bit resolution at 1 MHz, giving a full-frame readout time of $\sim$30s. The typical system noise for the CCD is given by the manufacturer as $\sim$9 $e^{-}$ at 1 MHz readout speed. We find the dark current to be $<$0.26 e$^{-}$ pixel$^{-1}$ s$^{-1}$ at a temperature of -20$^\circ$C and the CCD has temperature stability of $\pm$0.1$^\circ$C. The CCD operates at a gain of 1.4 $e^{-}$ ADU$^{-1}$. The CCD specifications list the full-well depth as $\sim$100000 e$^{-}$, but the analog-to-digital converter (ADC) saturates at 65535 ADU ($\sim$92000 $e^{-}$). The linear dynamic range is listed as 80 dB and the photoresponse non-linearity and non-uniformity are given as 1\%. The CCD also has anti-blooming protection to prevent image bleed from over-exposed regions.

\begin{figure}[ht]
	\begin{center}
		\plotone{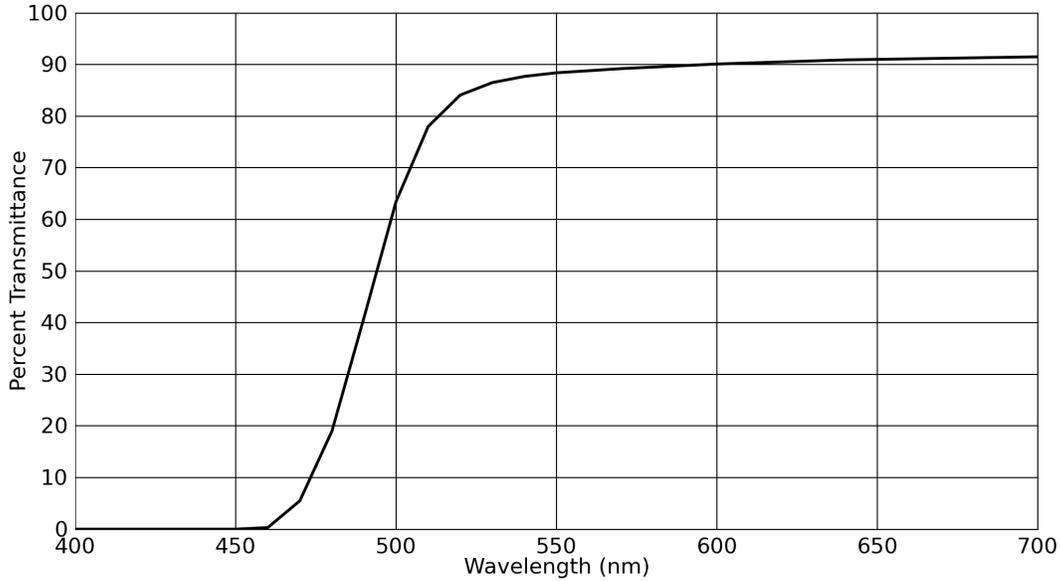}
		\caption{The theoretical transmission curve of the Kodak Wratten No. 8 Filter.}
		\label{fig:Wratten8}
	\end{center}
\end{figure}

KELT-South is equipped with a Mamiya 645 80mm f/1.9 medium-format manual focus lens with a 42mm aperture. The entire field of view using this lens is 26$^\circ$ $\times$ 26$^\circ$ and provides roughly 23'' pixel$^{-1}$ image scale. To improve star-sky contrast and minimize systematics, the KELT-South telescope uses a Kodak Wratten No. 8 red-pass filter with a 50\% transmission point at $\sim$490nm (See Figure ~\ref{fig:Wratten8}). The filter is mounted in front of the Mamiya lens.

\subsection{Mount}
\label{sec:Mount}
The optical assembly is mounted on a Paramount ME\footnote{http://www.bisque.com/sc/pages/Paramount-ME.aspx} robotic German Equatorial telescope mount manufactured by Software Bisque.  The German Equatorial configuration requires the telescope to perform a meridian flip when taking images on either side of the meridian.  This means that images taken east of the meridian are inverted and mirrored when compared to images of the same field taken west of the meridian.  The Paramount ME has integrated telescope and camera control but the camera control feature is not used in the KELT-South configuration.  Instead the camera is controlled directly through a USB cable connected to the control computer. The tracking error of the mount is $\pm$ 5''. This is much smaller than the large pixels of the CCD and does not affect our observations.

The mount has two predefined points that are used extensively during observing operations. The first is called the ``Home'' position and is defined by the manufacturer.  When the telescope is homed, the built-in circuitry on the axes of the mount knows the orientation of the mount. The second position is called the ``Park'' position and can be set by the user. For the KELT-South telescope the park position was chosen to be on the east of the pier facing north and slightly downward. This ensures that during idle periods the lens and filter are oriented downward such that dust accumulation is minimal.

\subsection{Observing Site}
\label{sec:ObsSite}
The telescope is located at the South African Astronomical Observatory site (32$^\circ$22'46" S, 20$^\circ$38'48" E, altitude 1760m) in Sutherland, South Africa. Sutherland is located in the western interior of South Africa about 370km to the north-east of Cape Town on the arid Karoo plateau.

Weather conditions at the Sutherland site are comparable to sites like La Silla and Paranal. Wind speeds less than 45 km/h occur 90\% of the time throughout the year and the median relative humidity level is 45\% (includes day and night time).  The median seeing conditions at the site is $\sim$0.92'' \citep{Erasmus:2000}. This means that 70\% of all nights are good for observations with 60\% of that time being measurably photometric. Since our point-spread functions (PSFs) are between 3 and 5 pixels, and the pixel scale is 23'' pixel$^{-1}$, atmospheric seeing variations are not a factor in the telescope observations.

\section{Observing Operations and Data Handling}\label{sec:obs_data}
The KELT-South telescope is completely robotic. No remote real-time observations are undertaken. The telescope is controlled by various VBScript programs that are executed by the Windows Scheduler at specific times. Observations are carried out every night with suitable weather conditions.

Observing operations start at 17:00 South African Standard Time (SAST) every day. The scripts first determine the current time using the Windows system clock and assess the weather conditions.  If any of the following are true, the weather is considered unsuitable for observing:

\begin{itemize}
\item The outside temperature is below -10 degrees C.
\item The humidity is above 90\%.
\item The difference between current exterior temperature and dew point temperature is less than two degrees (condensation is likely to occur).
\item The wind speed exceeds 60 km/h.
\item It is raining or snowing.
\item It is cloudy. 
\end{itemize}

All data from the active weather stations are averaged over the previous 15 minutes before making determination of bad weather. This is to ensure that sudden gust in wind speed or anomalous behavior of one weather station does not affect the overall procedure.

\subsection{Flat Fields} \label{sec:flats}

KELT-South takes two types of flat field images: dome flats and sky flats.  At the time of this writing, it is unclear which of these options will end up being best for final data reduction, so both types are currently acquired.  Dome flats are taken by pointing the telescope to the flat field screen on the underside of the roof of the enclosure.  The telescope points in a grid of sixteen positions around the center of the flat field, with the grid points offset by four arcminutes.  The dome flat exposures last two seconds.

Sky flats are taken each dusk and dawn.  At sunset, the control computer opens the roof and the telescope is pointed at the zenith. A test image is taken and the central 256 x 256 pixel block is read out and analyzed to determine the sky brightness level.  If the pixels are saturated, the telescope waits for five seconds and takes another test image.  That process continues until the sky brightness in the inner 256 x 256 pixel region is between 50,000 counts and 65,535 (ADC limit), with the goal being to acquire the most signal without hitting nonlinearity.  Then the telescope takes three full-frame images at 1-second exposures, making small ($\sim2$ arcmin) random adjustments to the position of the telescope between every image. Another test image is taken and the sky brightness level (modal value in the central image section) is determined again. If the sky brightness level has decreased, the exposure time is increased by one second for every 10,000-count drop in brightness, to ensure that all skyflat images have roughly the same number of counts.  Another three full-frame images are taken and the procedure repeated. The evening skyflat procedure stops once the exposure time required is over 13 seconds.  At this point stars start to appear in the images and further images cannot be used for flat fielding anymore. A similar procedure is done at dawn to take morning skyflats.

\subsection{Observing Procedures} \label{sec:obs}

If the scripts determine that conditions are not good for observing, a series of ten dome flats, five bias frames and five dark images are taken.  The telescope then enters a standby mode, in which it checks the weather conditions every five minutes, and if the weather conditions become good, observations start at that point.

If the weather is good, information is read into memory about Julian Day Number, Local Mean Sidereal Time at midnight SAST, rise and set times of the sun, start and end times of astronomical twilight, Local Sidereal Time at the end of astronomical twilight, the rise and set times of the Moon, illumination percentage of the Moon and also the Right Ascension and Declination of the Moon at midnight. At sunset, the skyflat procedure is performed (see above, \S \ref{sec:flats}).  After skyflats are taken, the roof closes up again, the telescope slews to the home position, and it takes ten dark images and ten bias images, in alternating pairs.  At the end of astronomical (18$^{\circ}$) twilight the telescope will start observations. Observations continue until astronomical dawn, at which point another set of skyflats are taken, the roof closes, the telescope is slewed to its park position, and an alternating set of five dark frames and five bias frames are taken, ending the observations for the night.

KELT-South has operated in both a commissioning mode and a survey mode.  In both modes all observations are taken with 150s exposures, selected to get optimal photometric precision of stars in the target range of $8 < V < 10$.  In the commissioning mode, the telescope observed a single predetermined target field for the entire night.  That mode was used to observe two different fields right after KELT-South deployment to serve as commissioning data.  The first field was centered on the open cluster Blanco 1, and the campaign on that field started September 16, 2009, and ended December 20, 2009, and comprises a total of 2,123 good-quality images from 43 separate nights.  The second field encompasses a number of open clusters, including NGC 2516, NGC 2547, and IC 2391, and the campaign on that field started January 3, 2010, and ended Feb 18, 2010, and comprises a total of $\sim3000$ images from 32 separate nights.  The cadence for commissioning mode observations is three minutes (150s exposures and $\sim30$s for readout).

In survey mode, the telescope observes a number of fields located around the sky throughout the night.  In that mode, the telescope first determines which fields are above 1.5 airmasses east of the meridian, and observes each field in order from Northernmost declination to Southernmost.  It then performs a meridian flip, determines which fields are available west of the meridian, and observes those fields in the same manner.  As targets set below the 1.5 airmass limit, new fields become available east of the meridian and the telescope will tile between available fields for the entire night. The telescope will not observe a field if it is within 45$^{\circ}$ (roughly two fields widths from field center) of the Moon.

There are two types of fields observed in survey mode.  Transit survey fields are observed every time the telescope cycles though the visible fields.  Although the cadence for transit survey fields depends on a number of factors, such as the number of fields visible at a given point in time and the position of the Moon, typical cadence for those fields is 10 to 20 observations per night.  Furthermore, two of the transit survey fields, which are located at declinations of +3$^\circ$ are being jointly observed with the KELT-North telescope.  In survey mode we also observe two fields at a different cadence, which are intended for use for stellar science investigations.  Those two fields are observed at most only two or three times per night, and are not expected to yield significant data for discovering transiting planets.

\begin{figure}
\plotone{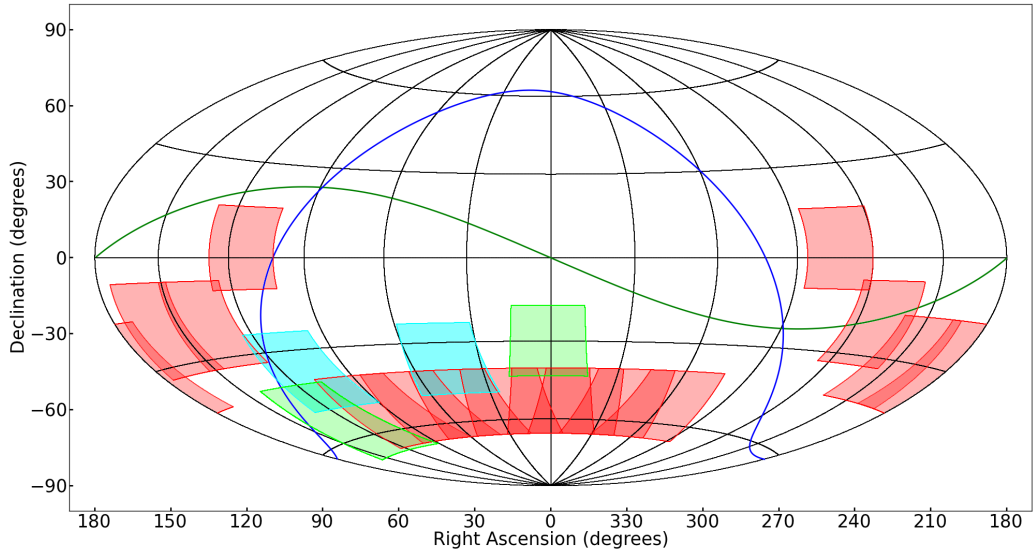}
\caption{Locations of the KELT-South observing fields.  Transit survey fields are in red, commissioning fields are in green, and stellar science fields are in cyan.  The commissioning field located at RA$=0^h$, Dec = -30$^\circ$, is also a transit survey field.  The blue line represents the Galactic Plane, and the green line indicates the ecliptic.}
\label{fig:fields}
\end{figure}

The current set of KELT fields are displayed in Figure \ref{fig:fields}, and the field centers are listed in Table \ref{tab:fields}.  The field locations were chosen through a combination of factors, including coordination with other surveys and the desire to avoid the Galactic Plane (due to concerns about crowding and the number of false positives).  That process yielded a set of fields located around the mid-latitudes of the southern hemisphere, covering $\sim45\%$ of the Southern sky.

In practice the typical cadence for a survey field is 15-20 minutes, since generally only $\sim4$ fields are visible at a given time, and most fields are not visible through the whole night.  Thus the true number of observations per night can vary widely.  However, even at the highest cadence and largest number of points per night, KELT-South does not acquire enough signal over a single transit to positively identify it.  We rely on phase-folding of full lightcurves over one or more years to build up the required signal for transit detection.  In that regime, the major factor in finding transits is not number of points acquired per night, but rather the total number of observations made.

\begin{table*}[t]
\centering
\scriptsize
\begin{tabular}{lcc}
Field Type & Right Ascension & Declination \\
 & (deg) & (deg) \\
\hline
transit survey  &   1.05  & -30.00 \\
\,\,\& commissioning &        &        \\
transit survey  &   1.05  & -53.00 \\
transit survey  &  22.95  & -53.00 \\
stellar science &  45.00  & -36.00 \\
transit survey  &  46.05  & -53.00 \\
transit survey  &  69.00  & -53.00 \\
transit survey  &  91.95  & -53.00 \\
stellar science & 109.95  & -37.00 \\
transit survey  & 115.05  &  +3.00 \\
\,\,\& joint with KELT-North &        &        \\
commissioning   & 124.00  & -54.00 \\
transit survey  & 138.00  & -20.00 \\
transit survey  & 160.95  & -20.00 \\
transit survey  & 184.05  & -30.00 \\
transit survey  & 207.00  & -30.00 \\
transit survey  & 229.95  & -20.00 \\
transit survey  & 253.85  &  +3.00 \\
\,\,\& joint with KELT-North &        &        \\
transit survey  & 298.95  & -53.00 \\
transit survey  & 322.05  & -53.00 \\
transit survey  & 345.00  & -53.00 \\
\hline
\end{tabular}
\caption{Locations of KELT-South observing fields}
\label{tab:fields}
\end{table*}

\subsection{Data Handling} \label{sec:Data_Hand_Arc}

Data acquired by the telescope throughout the night is stored on a hard drive in the control computer. In the morning, the data are compressed and copied to two external 200GB USB hard drives connected to the control computer. Using rice compression software (see \citet{Pence:2009,Pence:2010}, and references therein), the raw data are compressed by a factor of $\sim$40\%. 

When the drives reach 90\% capacity, they are manually swapped out for a new pair of drives, and the full drives are transported to SAAO in Cape Town.  One drive is shipped to Vanderbilt, and the other stays in Cape Town as a backup copy.  Once the shipped drive arrives in the U.S., and the data are copied from the drive and verified to be uncorrupted, the drive is shipped back to South Africa, and both drives are then returned to Sutherland.

The data transfer through shipped hard drives was necessary when KELT first began operating due to bandwidth limitations from South Africa.  However, major upgrades to the internet connection between South Africa and the rest of the world have recently been implemented, and expansion of the connection between Sutherland and Cape Town is also planned during the end of 2011.  We anticipate being able to transfer all KELT-South data directly to Vanderbilt from Sutherland by the first half of 2012.

\section{Telescope Performance}\label{sec:char}

The KELT project uses a heavily modified version of the ISIS\footnote{http://www2.iap.fr/users/alard/package.html} difference image analysis package \citep{Alard:1998,Alard:2000} to achieve high precision relative photometry.  The details of the KELT data pipeline will be described in an upcoming paper (Siverd, et al. in prep).  Here we present a broad outline of the pipeline, to provide the reader with enough context to evaluate the analysis of telescope performance in the following sections.

\subsection{KELT Pipeline Overview}\label{sec:pipeline}

First, individual images are dark-subtracted and then flat-fielded.  A "master" dark is produced for each observing night from a median combination of dark frames acquired at the beginning and end of nightly observing (typically $\sim15$ total).  All frames are acquired with the same exposure time (150s) as the science images.  There exists one master flat field that is used for all science images.  This image was carefully constructed by median combination of tens to hundreds of twilight sky flats, each of which was individually bias-subtracted, scaled-dark-subtracted, and additionally gradient-corrected prior to combination.  We are currently testing the stability of the use of a master flat in this manner for multi-season KELT data.

Calibrated images are distilled into lightcurves of individual objects using an image subtraction pipeline based on a heavily-modified version of the ISIS package.  All lightcurves for a field are derived over the total baseline of the observations, rather than in separate time segments such as per night or per season.  The exception to that is that because the design of our German Equatorial mount introduces a 180-degree rotation between images observed to the east and west of the meridian, we must reduce images for a given field in separate east and west sets, and then recombine the lightcurves at the end.  Also, any time the focus is adjusted, the images before and after the adjustment might need to be separately reduced, yielding pre- and post-focus adjustment lightcurves that are then combined.

As a first step, the pointing scatter is examined and a suitable high-quality image is identified near the median sky position of all available frames to serve as an astrometric reference.  Each individual image is then registered (aligned) to this image.  Once all images are registered to a common pixel grid, the highest-quality images are identified and combined to assemble a so-called reference image.  Relative to the overall image population, the reference-grade images are typically those that (a) were acquired at high altitude / low airmass, (b) have low sky background flux (i.e., taken when the Moon was set), (c) have lower full-width at half-maximum (FWHM) (i.e. no thin clouds or sky turbulence), and (d) exhibit high stellar flux.  These highest-quality images (up to 100 or 200 in number if possible) are then median-combined.  The resulting reference image is a maximally-high S/N image (in both flux and position) that we use to define the positions (both pixel and celestial) and fluxes of all objects identified for extraction. 

With an acceptable reference image, the pipeline begins in earnest.  The reference image is convolved to match the object shapes and fluxes for each individual image, and the convolved reference is then subtracted from that individual image.  By first matching object shapes in this fashion, we ensure that any residual flux has the same shape (PSF) as the original image.  The pipeline then measures the residual flux from the subtracted images using PSF-weighted aperture photometry.  Photometry is performed at a list of positions corresponding to objects identified on the reference frame.  These flux differences are then added to the median flux (obtained directly from the reference image using DAOPHOT II PSF photometry) in order to assemble individual lightcurves.  

\subsection{Pointing and Tracking}

We employ the TPoint software suite (integrated with TheSky6) to assist with telescope pointing. The current pointing model contains 200 points distributed evenly across both altitude and azimuth. The 200 points were collected on one night after the telescope was polar aligned using the methods set out by the TPoint software.  TPoint allows us to determine the polar misalignment of the telescope, which is currently misaligned by 29'' in azimuth and 22'' in altitude. This represents an overall error of less than 0.05\% of the entire size of the field. We use TPoint to determine that we have an overall pointing accuracy rms of 46.68'', which corresponds to roughly two pixels.  We also monitor the coordinates of the center of the KELT fields and find that over an entire observing season the typical pointing scatter in angular separation from expected central coordinate is 1.5 arcminutes, about 3.9 pixels.

\subsection{Astrometric Precision}

We acquire astrometric solutions for our reference images using the astrometry.net software package \citep{lang2010}.  Our astrometric solutions derived using this method are quite good.  We measure this by comparing our astrometrically derived coordinates to the celestial positions from the Tycho-2 catalog \citep{tycho} of a sample of stars across a reference image for one of the KELT commissioning fields.  We find that the median offset between our coordinates and Tycho coordinates is 7.9'' (about 1/3 of a pixel), with a standard deviation of 4.3''.  That astrometric accuracy applies to KELT observations away from the Galactic plane.  In the galactic plane, where extreme crowding reduces centroid accuracy and precision, we find a larger median offset of 17 +/- 8.4 arcseconds.

\section{Image Quality and Photometric Performance} \label{sec:image}

\begin{figure}
\includegraphics[scale=0.8,angle=0]{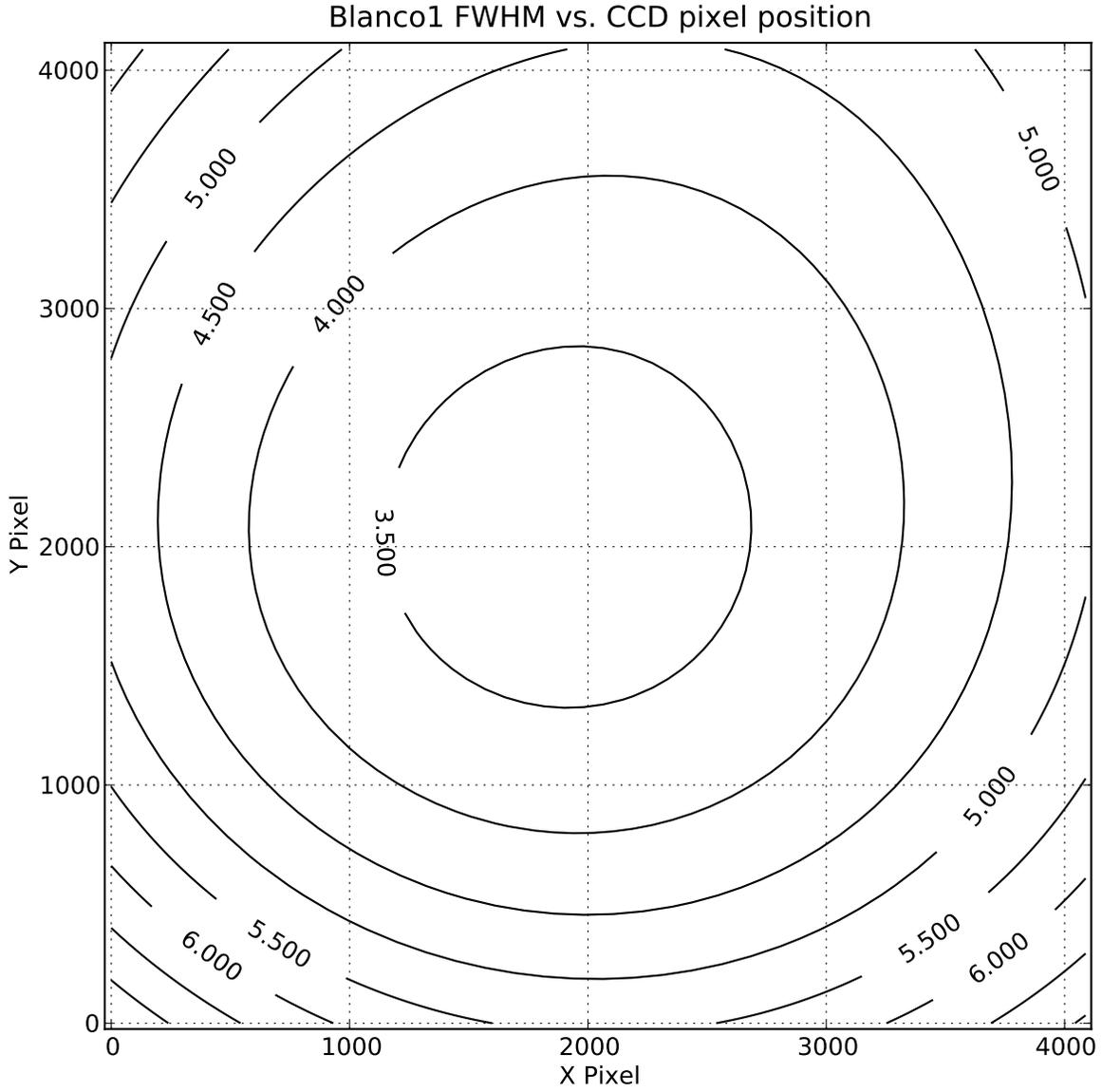}
\caption{Contour plot showing how the FWHM of the PSF varies across the KELT-South field.  Contour labels are in pixels.}
\label{fig:psf_contours}
\end{figure}

\begin{figure}
\includegraphics[scale=0.8,angle=0]{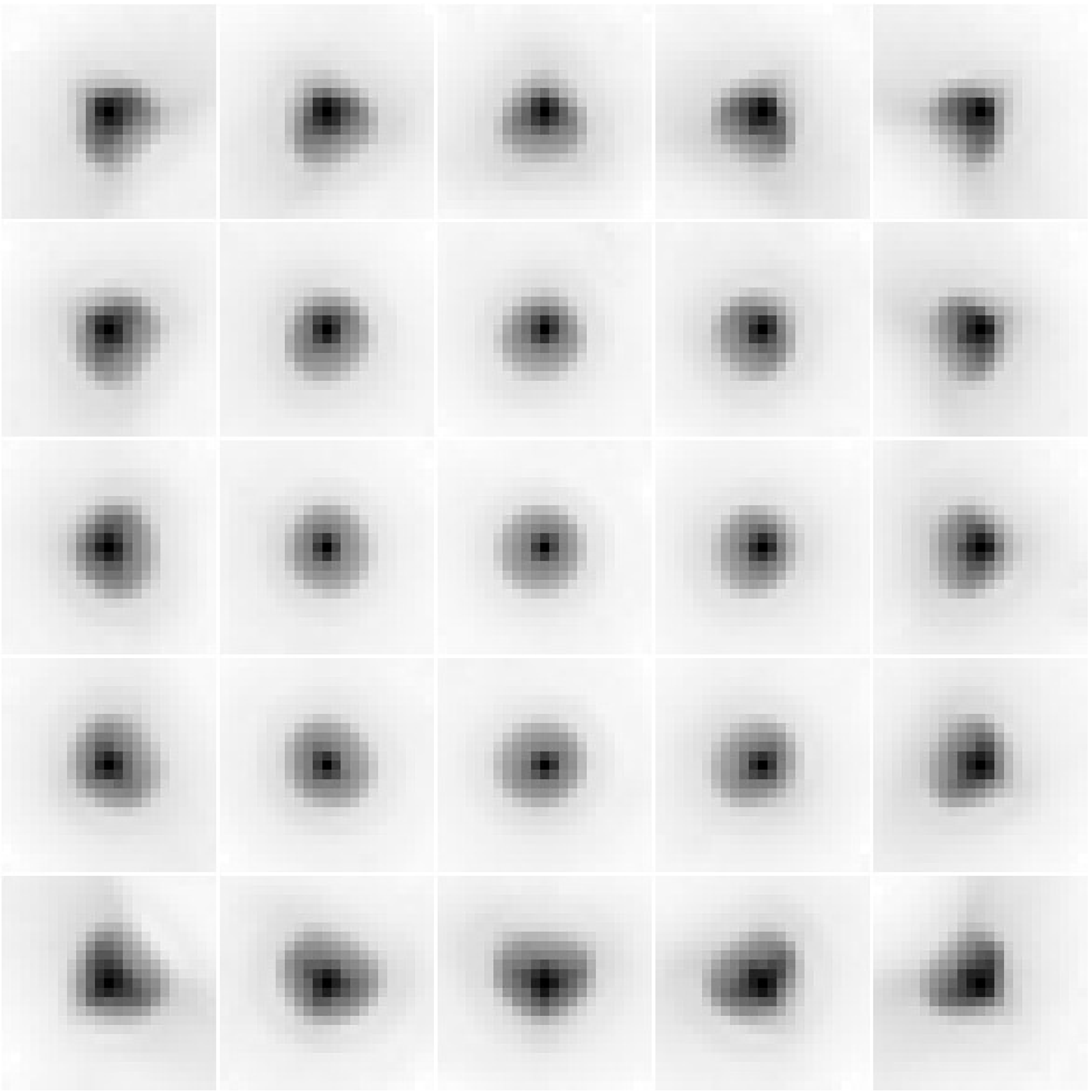}
\caption{Image stamps of bright, nonsaturated, isolated stars located in the respective regions of the KELT field.  This image shows how the shape of the PSF is circular at the field center and becomes distorted towards the edges and corners.}
\label{fig:psf_stamps}
\end{figure}

The wide field of view of KELT-South creates significant distortions of the PSF across the observing field.  Near the center of the field, the PSFs are round and have a FWHM of about 3 to 3.5 pixels.  Towards the edges and corners of the field the PSFs become distorted into a triangular shape, and grow to as large as 5 or 6 pixel FWHM.  There is also a decrease in flux of $\sim30\%$ between the center and edge of the field due to vignetting.  Figure \ref{fig:psf_contours} shows the distribution of PSF sizes across the field in a contour plot, and  Figure \ref{fig:psf_stamps} depicts the changing PSF in different regions of the image.  This PSF change contributes to the wide range in RMS magnitude scatter at fixed stellar magnitude, as described below in Section \ref{sec:relphot}. 

\subsection{Photometric Calibration}

Although KELT-South uses similar hardware to KELT-North, the camera is a newer model (see \S \ref{sec:Camera}), with a different gain.  For that reason, the conversion from KELT-South instrumental magnitudes to standard magnitudes is different.  We compared a set of nonsaturated KELT stars across an entire field (which incorporates any off-axis color terms) to the Johnson $V$ and Cousins $R$ magnitudes for those stars listed in the NOMAD catalog \citep{nomad}.  We find that KELT-South instrumental magnitudes correspond to $V$ and $R$ as follows:

\begin{equation}
V = KELT - 2.07
\end{equation}
\begin{equation}
R = KELT - 2.54
\end{equation}
where the calibration to $V$ has an average scatter of 0.22 magnitudes, and the calibration to $R$ has an average scatter of 0.28 magnitudes.

\subsection{Relative Photometry} \label{sec:relphot}

\begin{figure}
\includegraphics[scale=0.7,angle=0]{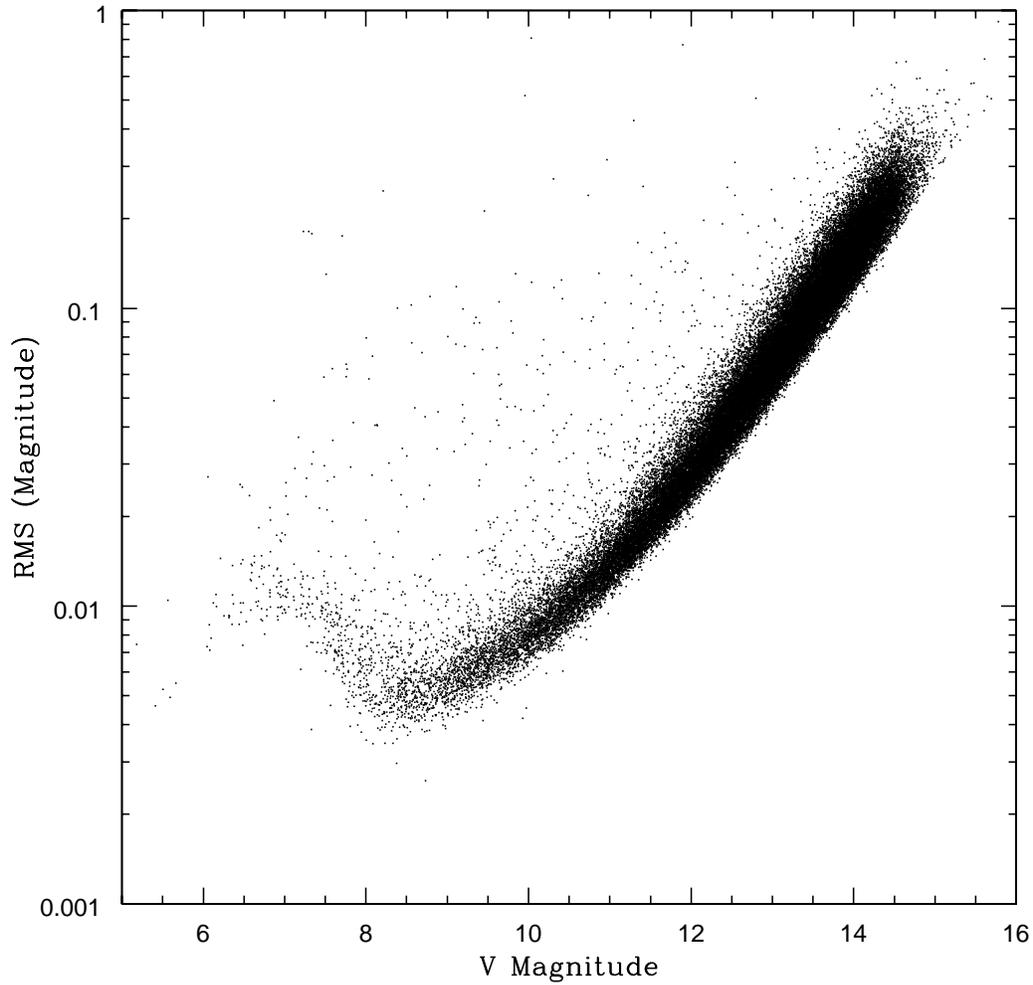}
\caption{ RMS plot of the first KELT-South commissioning field.  The field was observed for a duration of 96 days, with good data acquired on 43 separate nights.  Each lightcurve consists of up to 2,123 data points after sigma-clipping to remove outliers.  The upturn at the bright end indicates the onset of nonlinearity and saturation effects.}
\label{fig:rms_v}
\end{figure}

As described above (\S \ref{sec:pipeline}), KELT uses a difference image analysis pipeline to derive precise relative photometry.  There are several ways to measure the overall photometric precision of a data set, but the most common method is with an RMS plot.  Figure \ref{fig:rms_v} shows an RMS plot of the first KELT-South commissioning field (the one observing the field of Blanco 1).  The observations were taken from September 16, 2009, to December 20, 2009, and comprise a total of 2,123 images from 43 separate nights after elimination of poor-quality images.  After removing the three highest and lowest points for each lightcurve and performing one round of 7-sigma clipping to remove outlier data points on each lightcurve, we find that 3,792 stars have photometry with better than 1\% RMS, of which 3,564 are in the range $8 < V < 11$.

While the cadence used to observe the KELT commissioning fields was different from the typical survey fields, all other properties of the observations and data reduction are the same, and thus the results in Figure \ref{fig:rms_v} demonstrate the typical precision achievable by the KELT-South survey.  Note that those results do not include applications of the TFA detrending algorithm \citep{tfa}, which we find in initial testing increases the number of stars with sub-1\% photometry in a typical field by a factor of $\approx 20\%$.

\begin{figure}
\includegraphics[scale=0.5,angle=0]{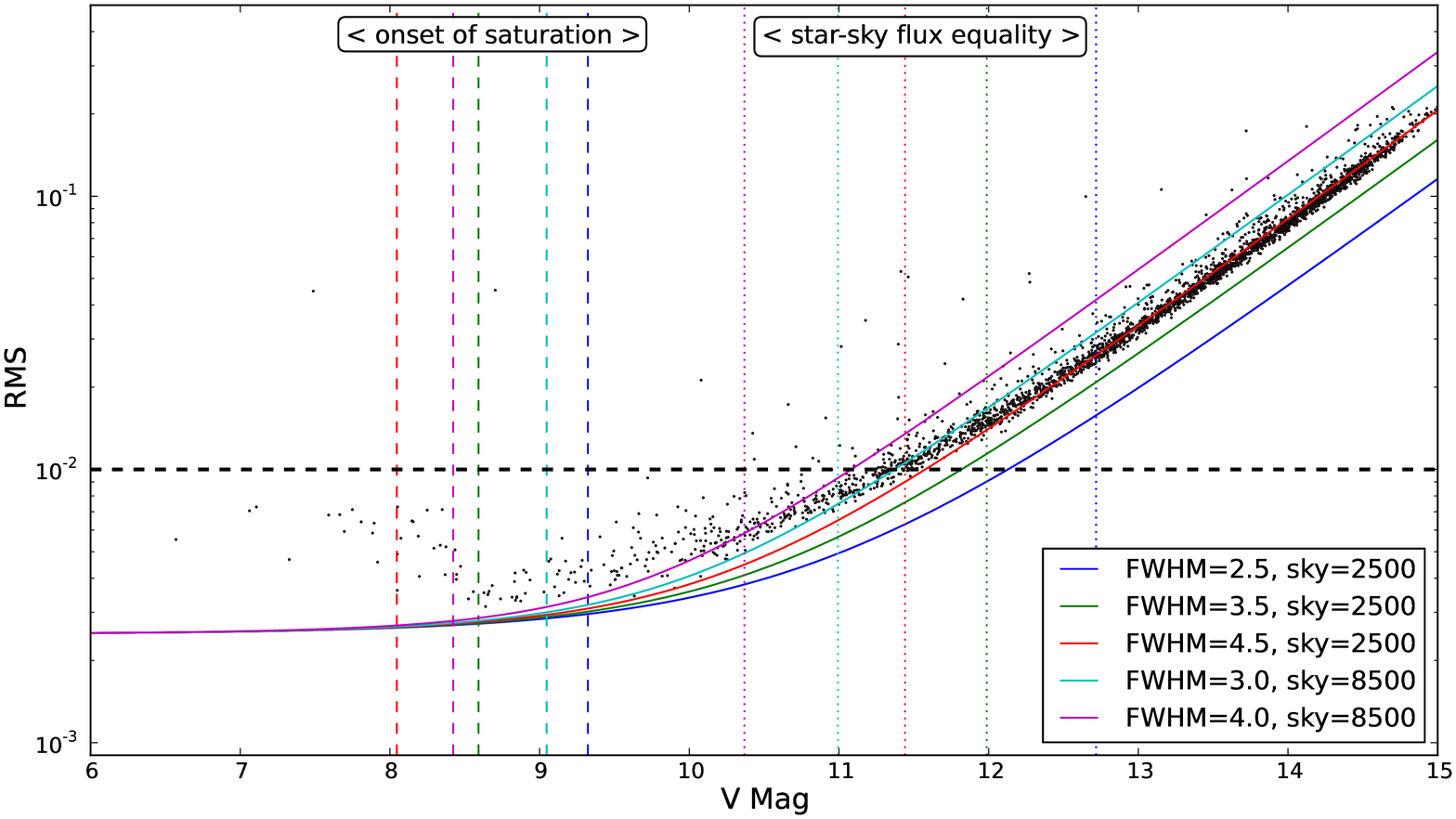}
\caption{RMS plot of the central portion of the first KELT-South commissioning field.  This plot includes many fewer stars, but displays the performance of KELT-South in greater detail.  The solid curves indicate theoretical photon noise limits, and the  dotted and dashed vertical lines respectively indicate the corresponding magnitudes of star-sky flux equality in a PSF, and the onset of ADU saturation in the brightest pixel.  The theoretical photon noise curves all include an assumed systematic noise floor of 2.5 mmag}
\label{fig:rms_d}
\end{figure}

Figure \ref{fig:rms_d} shows a detailed RMS plot, incorporating lightcurves from stars in the central 800 x 800 pixel section of the detector, and including TFA-based detrending of the lightcurves, using only data from the East orientation of the telescope.\footnote{Since Figure \ref{fig:rms_d} uses information about the size of the PSF, and the distribution of PSF is not exactly radially symmetric (see Figure \ref{fig:psf_contours}), the exact placement of the lines in Figure \ref{fig:rms_d} depends on what orientation the telescope is in.  The overall difference between the East and West orientation versions of this plot, however, is not significant.}  As Figure \ref{fig:psf_contours} demonstrates, the PSF size varies from the center to the edge of the detector.  Also, scattered light from Moon illumination can boost the background sky level substantially.  In Figure \ref{fig:rms_d}, we plot different curves showing the theoretical photon noise under different assumptions of PSF size and sky background.  Those assumptions yield different magnitudes for the point where the sky contributes an equal amount of counts to the star within a PSF, as well as the onset of ADU saturation in the brightest pixel, assuming a Gaussian-shaped PSF.  This Figure demonstrates that those two thresholds vary widely across the field and are time-dependent (i.e. based on Moon illumination), making it impossible to define a single value for either threshold.  However, the Figure shows that at the center of the field, well-behaved stars experience star-sky flux equality at roughly $V=11.5$ and start hitting saturation effect around $8 < V < 9$.  However, most stars that experience the onset of saturation still have sub-1\% RMS.

\subsection{Red Noise}

It is well-known that time-correlated systematic (red) noise can be a serious problem for transit surveys \citep{Pont:2006}, especially for small aperture, wide field surveys similar to KELT \citep{Smith:2006}.  Red noise can arise from a number of physical sources, such as temperature-dependent optical system changes, airmass effects, local weather conditions, etc.  Red noise can impede transit surveys by increasing the number of false positive candidates, and by obscuring real transit signals.  In order to characterize the level of red noise we see in KELT-South data, we compute the autocorrelation function on an intra-night time scale for a small number of bright, non-saturated, low-RMS stars observed in KELT-South commissioning data - stars representative of our best targets for transit detection.  

\begin{figure}
\includegraphics[scale=0.8,angle=0]{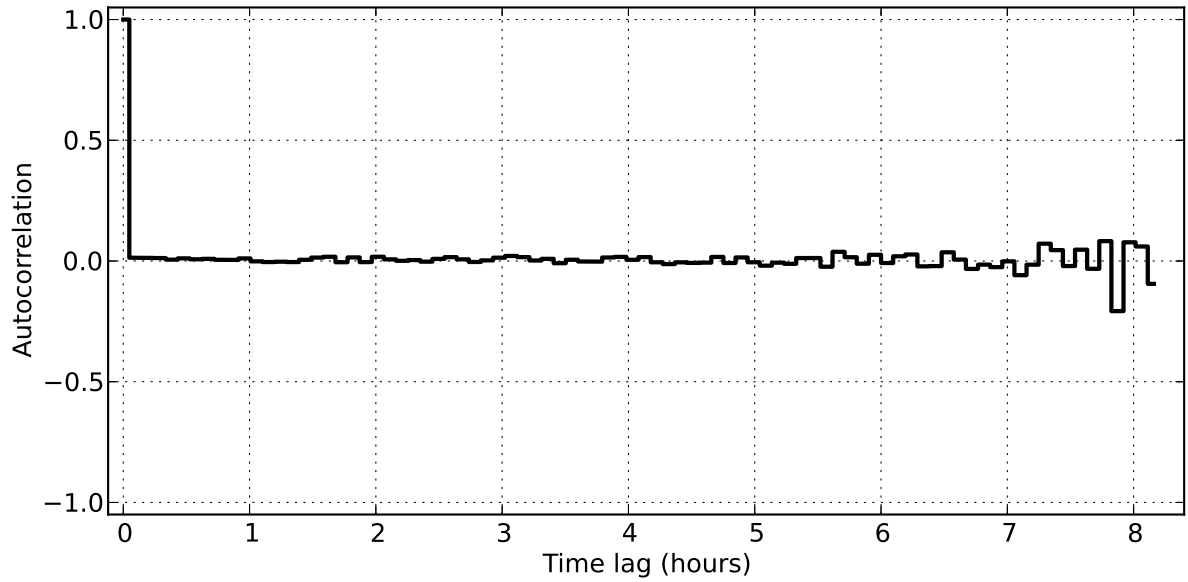}
\caption{Stacked autocorrelation functions of seven bright, non-saturated, low-RMS stars from KELT-South commissioning data.  No significant signal is seen on timescales of Hot Jupiter transits (2-3 hours).}
\label{fig:acorr}
\end{figure}

In Figure \ref{fig:acorr}, we plot the combined autocorrelation functions of seven bright stars over the timescale of one night.  As seen in the Figure, we do not see any evidence of correlated noise on timescales of 3 minutes up to the duration of a night ($\sim8$ hours).  The noise floor in the RMS plot in Figure \ref{fig:rms_d} might be due to some small amount of red noise, but at 2.5 mmag it is well below the expected 5 to 10 mmag depth of a giant planet transiting a solar-type star.  The combination of the low noise floor and the absence of red noise in the autocorrelation functions of the kinds of stars relevant to our transit search indicates that our ability to detect transits should not be compromised by large amounts of red noise.

\section{Sample Results}\label{sec:eg}

\begin{figure}
\includegraphics[scale=0.7,angle=0]{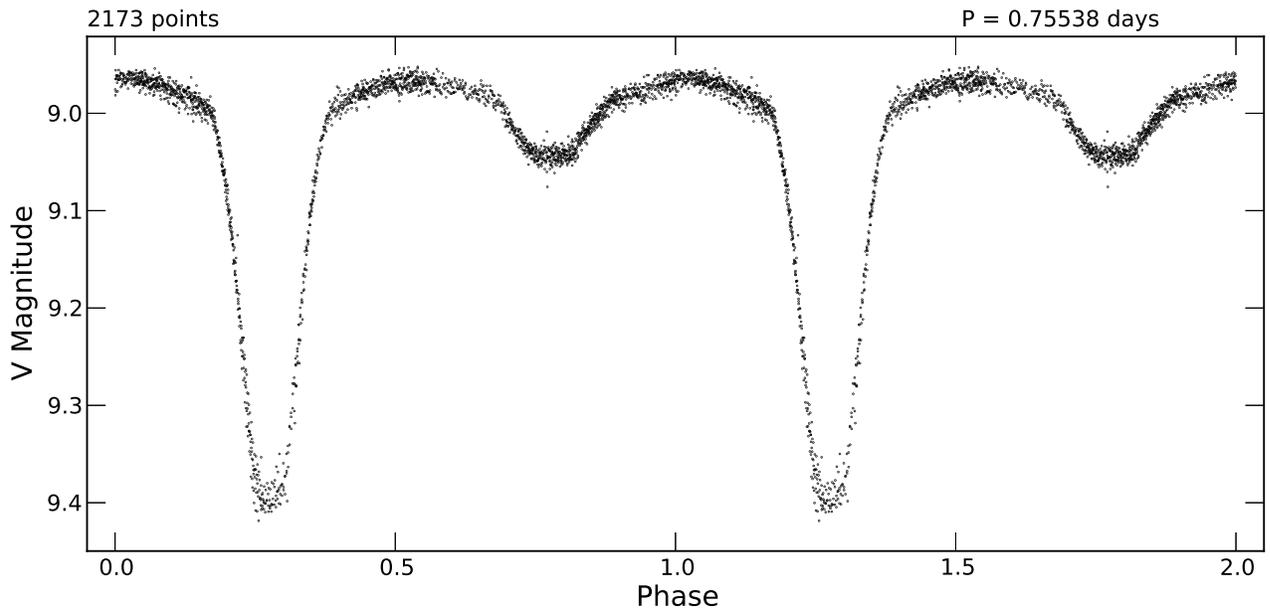}
\caption{Example lightcurve of eclipsing binary star AG Phe from KELT-South commissioning data.}
\label{fig:eb}
\end{figure}

\begin{figure}
\includegraphics[scale=0.7,angle=0]{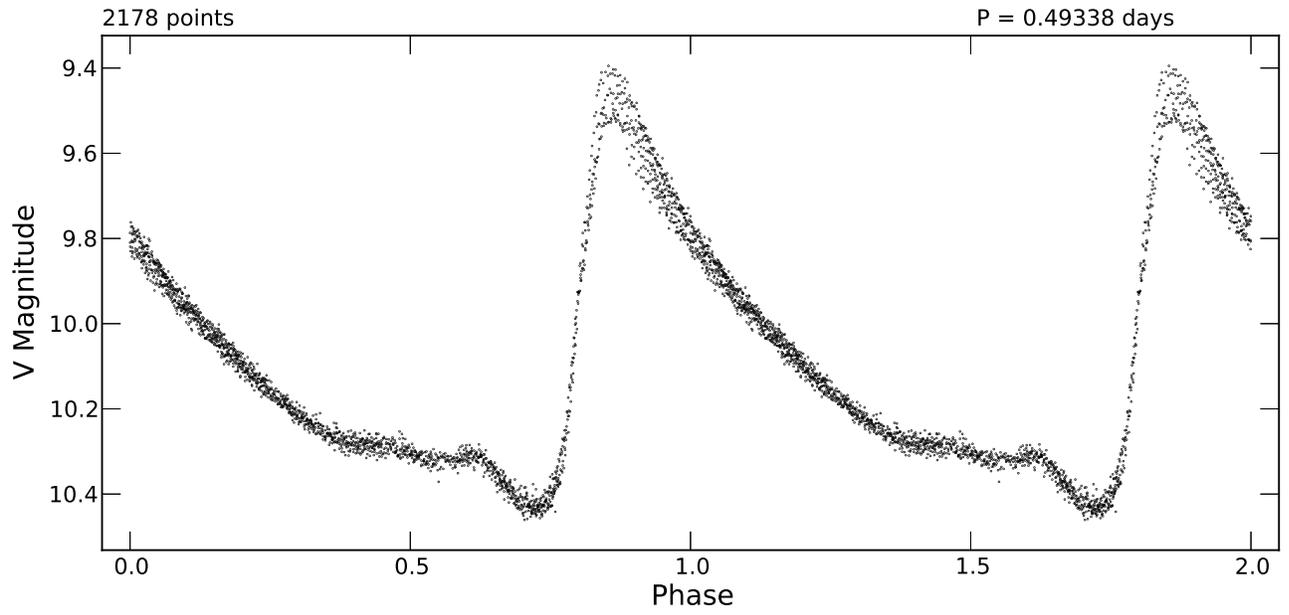}
\caption{Example lightcurve of RR Lyrae star RU Scl from KELT-South commissioning data.}
\label{fig:rrl}
\end{figure}

\begin{figure}
\includegraphics[scale=0.7,angle=0]{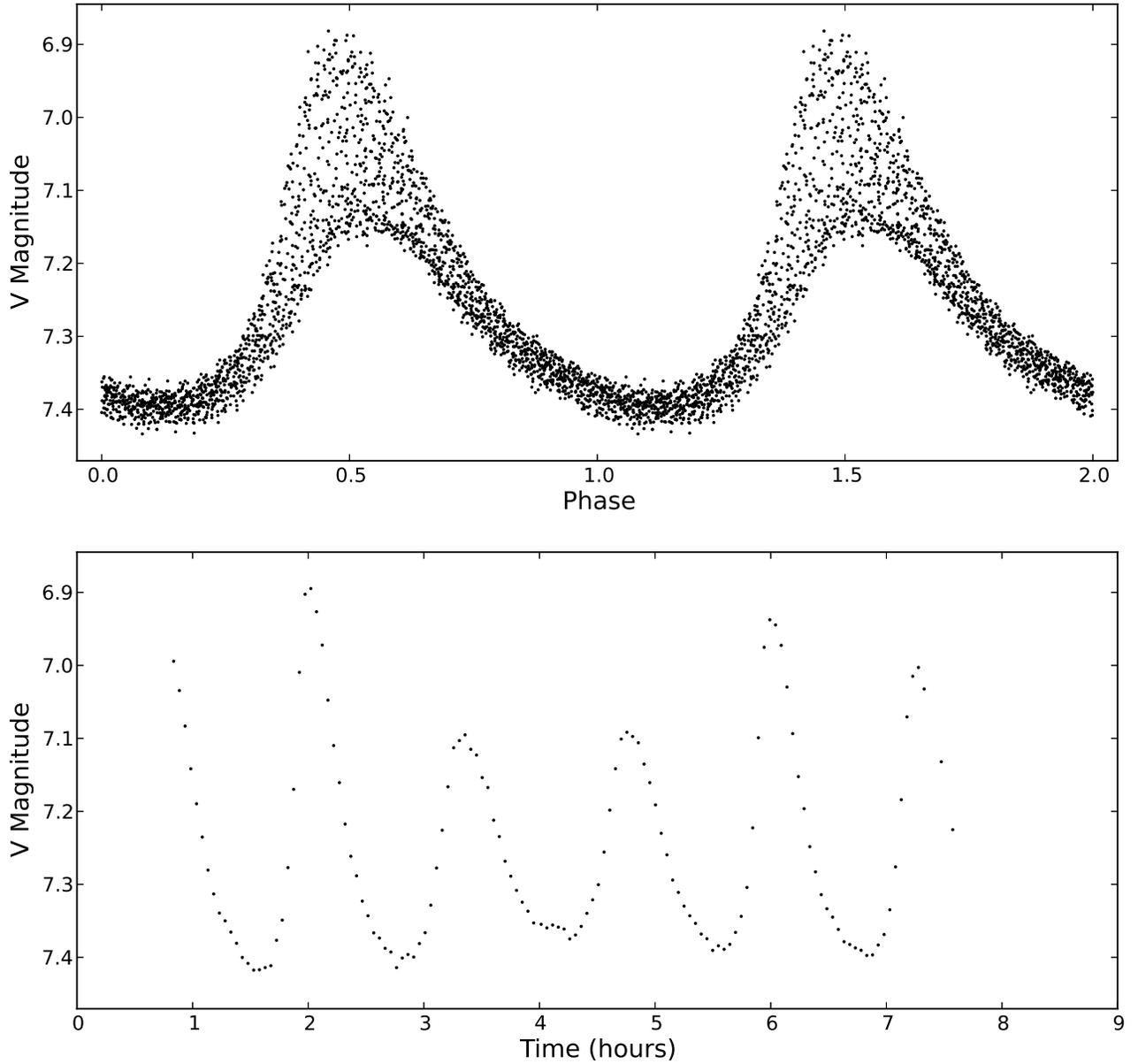}
\caption{Lightcurve of SX Phe from KELT-South commissioning data.  The upper panel shows the full phased lightcurve, at a period of 1.31923 hours.  The lower panel shows the data from one night of observations, displaying the rapid amplitude changes characteristic of this star.}
\label{fig:rcb}
\end{figure}

The plot in Figure \ref{fig:rms_v} demonstrates that KELT-South can obtain highly precise relative photometry over a timeline of many weeks.  To illustrate the potential value of KELT data for variable star science, we show sample KELT-South lightcurves of high-amplitude variable stars in Figures \ref{fig:eb}, \ref{fig:rrl}, and \ref{fig:rcb}.  Figure \ref{fig:eb} shows an Algol-type eclipsing binary star, AG Phe, with $V=8.9$, a period of 0.755385 days, and a depth of $\sim0.4$ mag.  Figure \ref{fig:rrl} shows an RR Lyra star, RU Scl, with $V\sim10$, a period of 0.493382 days, and an amplitude of $\sim1$ mag.  This target appears to display the Blazhko effect \citep{blaz}, but additional analysis will be needed to confirm that.  Figure \ref{fig:rrl} shows SX Phe, with $V\sim7.2$, a period of 0.493382 days, and an amplitude of $\sim1$ mag.  The upper plot shows the fully phased lightcurve with a 0.054968 day period, while the lower plot shows the data from a single night.

\begin{figure}
\includegraphics[scale=0.7,angle=0]{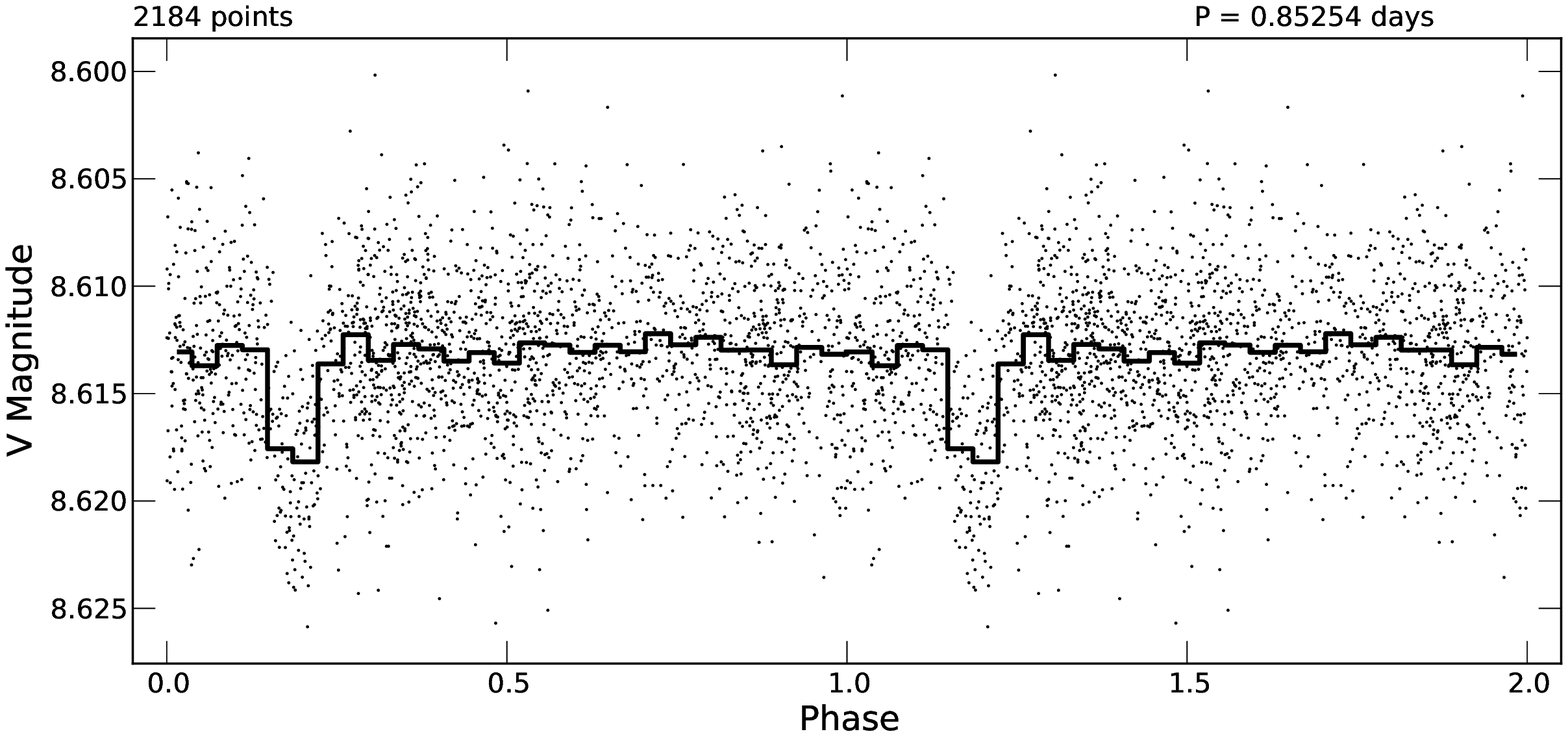}
\caption{Lightcurve from KELT-South commissioning data, showing a transit-like feature with a depth of $\sim5$ mmag and a period of 0.85254 days.  Black points are the TFA-detrended KELT data, and the solid line represents the same data binned at 38-minute bins.}
\label{fig:trans}
\end{figure}

To demonstrate the photometric precision of KELT-South and the potential for transit discovery with this set of data, Figure \ref{fig:trans} displays a transit-like lightcurve from KELT-South.  It shows a $V=8.6$ star with an RMS of 4 mmag, and a $\sim 0.5\%$ eclipse when phased to a period of 0.85254 days.  Although extensive followup will be required to determine whether this is a planet or one of the many common false positives, the lightcurve demonstrates our ability to detect signals similar to those of transiting planets.

\section{Summary}

In this paper, we have reviewed the properties and performance of the KELT-South telescope.  We have demonstrated that KELT-South can achieve the photometric precision needed to identify transiting giant planets.  This paper should also serve as a guide in proper use of KELT-South data.

KELT-South and KELT-North together are producing observations of time-varying sources across 42\% of the sky.  The main purpose of the survey is to identify bright transiting exoplanets that will be suitable for intensive followup, but we anticipate that the data will also be useful for identifying new variable stars and building comprehensive catalogs of specific objects in the magnitude range accessible to KELT.  We expect that such results will be of use to other future variability surveys for the purpose of preparing target lists and finding long baseline variability.  The examples shown in Section \ref{sec:eg} provide an indication of the precision and nature of the KELT data.  Finally, we intend to release all KELT lightcurves to a publicly accessible archive.

\acknowledgments 
We would like to thank the many individuals who assisted in all aspects of this project, including John Fellenstein, Bob Patchin, Nathan De Lee, Piet Fourie, Michael Rust, John Stoffels, Jaci Cloete, Willie Koorts, and John Menzies.  KELT-South received funding from the Vanderbilt Initiative in Data-Intensive Astrophysics (VIDA), FISK-Vanderbilt NSF PARRE grant AST-0849736, and the Vanderbilt International Office.

\newpage
\bibliography{ms}
\end{document}